\documentstyle[11pt, aaspp4]{article}
\begin{document}
\title{Cometary Dust in the Debris Disks of HD 31648 and HD 163296: Two ``Baby'' $\beta$ Pics}
\author{Michael L. Sitko}
\affil{Physics Department, University of Cincinnati, Cincinnati OH 45221}
\author{Carol A. Grady}
\affil{Eureka Scientific, 2452 Delmer St., Suite 100, Oakland CA 94602}
\author{David K. Lynch and Ray W. Russell}
\affil{The Aerospace Corporation, Los Angeles, CA 90009}
\author{Martha S. Hanner}
\affil{Jet Propulsion Laboratory, California Institute of Technology, Pasadena, CA 91109}

\begin{abstract}

The debris disks surrounding the pre-main sequence stars HD 31648 and HD 163296 were observed spectroscopically between 3 and 14 $\mu$m. Both possess a silicate emission feature at 10 $\mu$m which resembles that of the star $\beta$ Pictori and those observed in solar system comets. The structure of the  band is consistent with a mixture of olivine and pyroxene material, plus an underlying continuum of unspecified origin. The similarity in both size and structure of the silicate band suggests that the material in these systems had a processing history similar to that in our own solar system prior to the time that the grains were incorporated into comets.
\end{abstract}


\section {Introduction}

It is now known that a large fraction of all main sequence stars, perhaps most, possess significant amounts of refractory materials long after their formation (Aumann 1988). Originally discovered through their excess infrared emission, this ``Vega phenomenon'' has spurred many recent observational investigations into the formation and evolution of protostellar disks similar to that which our own sun once had, and which still remains in the form of interplanetary debris (dust, comets, etc.).

The star $\beta$ Pictoris is the best-studied example of a main sequence star with an extrasolar debris disk. The disk, which is viewed edge-on, extends hundreds of AU, and has a central clearing and warpage that may signal the presence of one or more jovian planets. The colors of the disk are relatively neutral at visible wavelengths (Paresce \& Burrows 1987) suggesting that the grains must be generally larger than 1 $\mu$m. The disk of $\beta$ Pic is also polarized at about 17\% (Gledhill et al. 1991), similar to the zodiacal light of our own solar system (Leinert et al. 1982), suggesting that the dust in the $\beta$ Pic disk includes grains similar to solar system interplanetary dust particles, including large (say 10 $\mu$m diameter) fluffy aggregates of submicron size components. Thus in may ways, the dusty debris disk in $\beta$ Pic is similar to that in our solar system, but more massive and presumably younger. Unlike the zodiacal light in the solar system where the band/continuum ratio is about 1.15 (Reach et al. 1996), $\beta$ Pic exhibits a silicate emission feature with a band/continuum ratio of about 2 (Knacke et al. 1993) that requires a significant population of grains smaller than a few $\mu$m. Since such small grains will be easily removed by radiation pressure, they must be contunuously replenished by the erosion of larger bodies. The emission of these small grains provides a means by which the mineralogy of the debris disk may be studied.

The shape of the IR spectrum of $\beta$ Pic (Knacke et al. 1993) is similar to that of some solar system comets in the sense that both have silicate emission features at 10 $\mu$m, including a subfeature at 11.2 $\mu$m due to crystalline olivine. Furthermore, Bradley et al. (1992) have identified the same spectral signature in a subclass of interplanetary dust particles that consist of porous chondritic glassy silicates with embedded crystalline material. Some of the subcomponents of these grains consist of glasses with embedded metal and sulfides (``GEMS'') whose physical and chemical structure suggests extensive cosmic ray exposure in the interstellar medium or in the protostellar nebula (Bradley 1994a, 1994b). Thus the debris disk of $\beta$ Pic and the solar nebula (and, by inference other protostellar disks) exhibit many of the same spectral characteristics and may have undergone similar evolutionary processes. To understand better how $\beta$ Pic and other main sequence stars with debris disks evolved into their present state, we need to investigate their evolutionary precursors.

The immediate evolutionary precursors of the the Vega-like stars
are the T Tauri stars for solar-mass stars and Herbig Ae/Be (HAEBE) stars for more massive stars like $\beta$ Pic. Classic HAEBEs contain emission lines in their spectra, lie in regions of star formation that include ``obvious'' oscuration due to dust, and illuminate nearby reflection nebulae (Herbig 1960). They also possess infrared excesses longward of 1 $\mu$m. However, it has become apparent that  there are a number of sources that have all the spectral characteristics of the classic HAEBEs, but are not embedded in dusty regions. Many are probably intermediate-age objects, not yet main sequence stars, but generally older than the classical HAEBEs: ``post-HAEBEs'' (PHAEBEs). Many exhibit evidence of sporadic accreting material (Grady et al. 1996) similar to that of $\beta$ Pic, and in one case, HD 100546, the data are consistent with stargrazing comets (Grady et al. 1997).

Mannings \& Sargent (1996) and Mannings, Koerner \& Sargent (1997) have mapped the thermal dust mm-continuum and gaseous CO emission in HD 31648 and HD 163296, two stars that would be classified as PHAEBEs. For HD 163296, deconvolution of the maps of the CO emission indicates a disklike structure which has a radial extent of 310 AU. The dust has a radial extent of 110 AU (HWHM). For HD 31648, the dust disk has an extent of 85 AU (HWHM). Both objects have accurate distances from the HIPPARCOS database, with HD 31648 at 131 pc and HD 163296 at 122 pc (both are uncertain by about 20 pc). These disks are much more massive than that of $\beta$ Pic and the other Vega-like main sequence stars, and may provide a glimpse of what they and our own solar system may have been like during their later stages of formation.

\section{Observations}

IR Spectra of HD 31648 and HD 163296 were obtained with the Aerospace Corporation's Broadband Array Spectrometer System (BASS) configured to f/35 at the 3m NASA Infrared Telescope Facility (IRTF). The BASS consists of a pair of cooled prisms which disperses the spectrum onto two 58-element blocked impurity band (BIB) linear arrays that simultaneously span the 3-13.5 $\mu$m region. The spectral dispersion ranges from about 30 to 125 over each of the 3-6 and 6.5-13.5 $\mu$m regions.  At the IRTF, the BASS entrance aperture subtends 3.4 arcsec. The observations reported here were obtained on 14 October 1996 (UT), along with observations of Comet Hale-Bopp as part of a group of long-term programs to observe the spectra of pre-main sequence stars and solar system comets. The spectra were flux-calibrated using observations of standard stars at nearly the same airmass and time. For HD 163296 and Hale-Bopp, observations of Vega (which has no detected excess emission in the wavelength region covered here) were used. For HD 31648, Arcturus was used. Details of the observations are listed in Table 1, and the resultant spectra of the two program stars are shown in Figure 1, where we have plotted the spectral flux ($\lambda$F$_{\lambda}$ in W m$^{-2}$) of the program stars. The BASS spectra of Comet Hale-Bopp are discussed elsewhere (Hanner et al. 1998; Russell et al. 1998).

It is quite apparent that the spectra of these two stars are quite similar. Both objects possess emission by warm dust down to 3 $\mu$m, and a well-defined silicate emission band near 10 $\mu$m, features that are common in many pre-main sequence stars accompanied by dust. In the following section, we shall examine how the data on these two stars compare to that of $\beta$ Pic, solar system comets, and laboratory spectra of relevant materials.

\section{Comparison with $\beta$ Pic, Solar System Comets, and Laboratory Measurements}

$\beta$ Pic has an emission band that rises up to a local maximum near 9.5 $\mu$m, is nearly flat-toppped to 11.2 $\mu$, then falls off to the local continuum (Knacke et al. 1993). This sort of ``trapazoidal'' silicate emission is also present in the spectra of many, but not all, solar system comets (Hanner, Lynch, \& Russell 1994). When normalized in a similar fashion (see for example Figure 6 of Knacke at al.) the short-wavelength peak appears somewhat higher in $\beta$ Pic than in the comets.

The feature near 11.2 $\mu$m deserves special attention because it is indicative of the presence of crystalline olivine. This feature is also seen in the ouflows of some evolved stars where dust is condensing (Waelkens et al. 1996), but is not a general feature of the dust in either star-forming regions no diffuse interstellar clouds. If we see it in comets and $\beta$ Pic, where is it made?

Since crystalline olivine condenses at the high temperatures ($>$1200 K) that might have been typical of the inner solar nebula, it could, in principle, have formed by condensation during the formation of the solar system. This scenario requires that these grains then be transported to the region of comet formation and be mixed with ice grains that probably condensed at temperatures below 30 K (Crovisier et al. 1997). How this is accomplished is not fully understood. Similarly, more amorphous grains may have undergone annealing in the solar nebula, but again would need to be efficiently transported out to the region of comet formation. In any case, such grains do exist in solar system comets, and examining the development of these features in other stellar systems may give us clues as to how it was accomplished in our own solar system.

Until recently, the evidence for comet-like silicate emission features in other pre-main sequence stars was scarce. Using data primarily from the IRAS Low Resolution Spectrometer database, the majority of HAEBEs and PHAEBEs exhibit little spectral structure in the silicate band (see Sitko et al. 1994, for example). However, in the past two years, a considerable mass of evidence clearly indicates that this conclusion may have been biased by studying very young objects, whose dust has undergone little physical and chemical processing since the disk formed. The intermediate-age objects appear to exhibit spectral structure that may correlate with the degree of disk clearing (van den Ancker 1997). HD 31648 and HD 163296 are the first of these intermediate age higher-mass stars, ``baby $\beta$ Pics'', to have resolved protostellar disks in both gas and dust emission, and both exhibit structure in their silicate bands similar to that of $\beta$ Pic.

In order to compare the spectral characteristics of the young stellar debris disks with that of solar system dust, it is useful compare the net emissivities of the dust in the silicate band region. As a first approximation, this can be done by dividing the observed flux by a gray body normalized to the continuum flux outside the emission band, as has been done in the past for cometary spectra (Hanner et al. 1994).  In Figure 2, the spectra of HD 31648 and HD 163296 have been normalized to a gray body between 8 and 13.5 $\mu$m . Also shown are the emission spectra of Comet Hale-Bopp (C/1995 O1), obtained within 30 minutes of the HD 163296 spectrum (and at the same airmass), and Comet Levy (C/1991 L3; Lynch et al. 1992) treated in the same fashion.

All four spectra possess similar characteristic features. They all have a general trapazoidal shape, reaching local maxima near 9.5 and 11.2 $\mu$m, and are all relatively flat-topped between those wavelengths. The band/continuum ratio in this region is between 1.5 and 2.5 for all four objects. The silicate feature in HD 31648 has both the same strength and shape as Comet Levy from 10 to 13 $\mu$m, and differs only in having a slightly greater emission at the short-wavelength wing. Similarly, the band emission in HD 163296 also shows excess emission at 9.5 $\mu$m compared to comets, as shown in Figure 3, where it is compared to the profile of Hale-Bopp after removal of the underlying continuum (as was done for $\beta$ Pic in Figure 6 of Knacke et al.). While the 9.5 $\mu$m feature sits in the middle of the telluric ozone band and is subject to greater uncertainty than most of the rest of the emission profile, we believe this difference is real. First, the error bars shown are simple standard deviations, which tend to overestimate the true uncertainty of the mean value (the standard deviation of the mean). More importantly, such a difference can arise from systematically having different ozone opacity for the HAEBES compared to their calibrators, but this is unlikely to be the source of the difference seen here for two reasons. First, in the case of HD 163296 the data for the star and Hale-Bopp were obtained through nearly the same airmass, at about the same time, and calibrated with respect to the same standard star observation. Second, {\it the peak at 9.5 $\mu$m is seen in many of these objects} (Sitko et al. 1999). Furthermore, the same difference seems to be present in the spectrum of $\beta$ Pic. 

It is instructive to compare the spectra of the dust in these systems to those of silicate materials measured in the laboratory. Stephens \& Russell (1979) have measured the relative emissivity of pyroxene and olivine in both glassy and crystalline forms. The glassy material was obtained by laser-vaporization of mineral samples. The resultant condensate smoke was allowed to settle on a copper block. Crystalline samples were prepared by grinding the mineral material and allowing them to be deposited on a copper block. The emissivity was then determined by heating the block plus sample and measuring the resultant spectrum. The material used was Mg-rich olivine and enstatite (specifically (Fe$_{0.07}$Mg$_{0.93}$)$_{2}$SiO$_{4}$ and  (Ca$_{0.01}$Fe$_{0.11}$Mg$_{0.88}$)SiO$_{3}$, respectvely). The ground particles were generally less than 5 $\mu$m in size, while the median diameter of the condensed glassy smokes was about 0.02 $\mu$m. The resulting emissivities of these four materials are shown in Figure 4.

In Figure 5, we compare the emission spectrum of HD 163296 with that of a simple model consisting of glassy enstatite, glassy olivine, and crystalline olivine with these emissivities. To this we have added a continuum source with an emissivity of unity (a blackbody). For simplicity, the same temperature, 440K, was assumed for all components. Crystalline olivine is required to fit the 11.2 $\mu$m feature, and glassy enstatite is needed for the short-wavelength wing of the band. The continuum source with T=440K is required to provide the underlying continuum flux and its spectral slope. The source of the continuum could be virtually any featureless material. Finally, some other component is needed for the middle of the band (crystaline enstatite could also work here, although it is a much narrower component). Given the simplistic nature of this model (using a few end-types of dust materials), the emission observed in HD 163296 is reproduced reasonably well by the model. However, the observed emission in this object (and by inference other HAEBEs with similar spectra, such as HD 31648) probably requires an additional component to explain the additional emission observed near 9.4 $\mu$m. It is likely that this is some form of pyroxene material. While we have restricted our fit to only one specific material, pyroxenes come in a wide variety of forms. While the infrared spectra of the olivine-domnated interplanetary dust particles studied by Sanford \& Walker (1985) all had similar spectra, no two pyroxene IDPs were identical!

\section{Discussion}

Overall, the spectral emission of these stars resemble those of ``typical'' solar system comets, suggesting that the dust in these systems may have undergone a history of processing similar to that of our own solar system prior to when the dust material was encorporated into comets like Halley and Levy. The majority of comets with well-defined silicate emission bands are long-period objects (Halley, though strictly short-period, has an inclination more like long-period objects) and presumably arise from the Oort cloud. In turn, these objects are believed to have arisen in the Uranus-Neptune region before being scattered outward (and inward) by planetary perturbations. The silicate band in the short-period comets, whose origin is in the Kuiper belt, is smaller than that seen in HD 31648 and HD 163296, having band/continuum ratios of only 1.25 or less (Hanner et al. 1996). 

In HD 31648 and HD 163296, we are presumably seeing the emission of some ensemble of cometary objects from the same region that gave rise to the Oort cloud comets in our own solar system. The ratio of luminosities of the dust emission of HD 163296 to that of Hale-Bopp, using the 4.3 arcsec aperture, between 8 and 13 $\mu$m, is 3x10$^{14}$. Thermal infrared images of Hale-Bopp (Hayward \& Hanner 1997) indicate that a much larger aperture would reduce this ratio by an order of magnitude, to a few times 10$^{13}$. A similar ratio was obtained by Malfait et al. 1998 for the PHAEBE star HD 100546 with respect to  Hale-Bopp using Infrared Space Observatory's Short Wavelength Spectrometer (ISO SWS) for both objects. (The ISO SWS entrance aperture is 8x20 arcsec.) Whether we are actually seeing the silicate emission from the comae of this many objects in HD 163296, or the emission from just a few large (Chiron-sized) disrupted objects, is not known. If it is the latter, it should be possible to detect changes in the band strength and shape over time scales of years or less, possibly accompanied by flux variability at visible and ultraviolet wavelengths.

Like $\beta$ Pic, many of the HAEBEs and PHAEBEs exhibit optical and ultraviolet spectral features indicative of infalling evaporating objects, such as comets or asteroids (Grady et al. 1996, 1997; P\'erez \& Grady 1997) some of which require star-grazing orbits. HD 163296 exhibits such infall events (Grady et al. 1999), but these are not observed in HD 31648, presumably due to its higher disk inclination (30 degrees versus 58 degrees for HD 163296, where 0 degrees is pole-on; Mannings \& Sargent 1996; Mannings, Koerner \& Sargent 1997). Such orbits may require the presence of gravitationally pertubing bodies, such as Jovian planets. Mannings \& Sargent have estimated the ages of HD 31648 and HD 163296 as 6 Myr and 5 Myr, respectively. In the case of the solar nebula, this would correspond to well after the formation of planetesimals and even Jupiters (Wetherill \& Stewart 1993; Cameron 1995; Pollack et al 1996). Such star-grazing objects are now routinely detected in our own solar system using satellites with coronagraphs (such as SOHO), and must have been much more common during the more lively early bombardment period of the solar system. Since we are presumably observing HD 31648 and HD 163296 during a similar epoch, such infall events, and the production of cometary silicate emission features, is expected. 

\clearpage

\begin{references}

\reference {} Aumann, H.H. 1988, AJ, {\bf 96}, 1415.

\reference {} Bradley, J.P. 1994a, Science, {\bf 265}, 925.

\reference {} Bradley, J.P. 1994b, Geochim. et Cosmochim. Acta, {\bf 58}, 2123.

\reference {} Bradley, J.P., Humecki, H.J., \& Germani, M.S. 1992, \apj,
{\bf 394}, 643.

\reference {} Cameron, A.G.W. 1995, Meteoritics, {\bf 30}, 133.

\reference {} Crovisier, J., Leech, K., Bockel\'een, D., Brooke, T.Y., Hanner, M.S., Altieri, B., Keller, H.U., and Lellouch, E. 1997, Science, {\bf 275}, 1904.

\reference {} Gledhill, T.M., Scarrott, S.M., \& Wolstencroft, R.D. 1991, MNRAS, {\bf 252}, 50P.

\reference {} Grady, C.A., P\'erez, M.R., Talavera, A., Bjorkman, K.S., de Winter, D., Th\'e, P.S., Molster, F.J., van den Ancker, M.E., Sitko, M.L., Morrison, N.D., Beaver, M.L., McCollum, B., and Castelaz, M.W. 1996, A\&AS, {\bf 120}, 157.

\reference {} Grady, C.A., Sitko, M.L., Bjorkman, K.S., P\'erez, M.R., Lynch, D.K., Russell, R.W., \& Hanner, M.S. 1997, \apj, {\bf 483}, 449.

\reference {} Grady, C.A. et al. 1999, in preparation.

\reference {} Hanner, M.S., Gehrz, R.D., Harker, D.E., Hayward, T.L., Lynch, D.K., Mason, C.G., Russell, R.W., Williams, D.M., Wooden, D.H., \& Woodward, C.E. 1998, Earth, Moon and Planets, in press.

\reference {} Herbig, G.H. 1960, \apjs, {\bf 4}, 337.

\reference {} Hayward, T.L., \& Hanner, M.S. 1997, Nature, {\bf 275}, 1907.

\reference {} Knacke, R.F., Fajardo-Acosta, S.B., Telesco, C.M., Hackwell, J.A., Lynch, D.K., and Russell, R.W. 1993, \apj, {\bf 418}, 440.

\reference {} Leinert, C., Richter, I., Pitz, E., \& Hanner, M.S. 1982, A\&A, {\bf 110}, 355.

\reference {} Lynch, D.K., Russell, R.W., Hackwell, J.A., Hanner, M.S., \& Hammel, H.B. 1992, Icarus, {\bf 100}, 197.

\reference {} Malfait, K., Waelkens, C., Waters, L.B.F.M, Vandenbussche, B., Huygen, E., \& de GRaauw, M.S. 1998, A\&A, {\bf 332}, L25.

\reference {} Mannings, V., Koerner, D.W., \& Sargent, A.I. 1997, Nature, {\bf 388}, 127.

\reference {} Mannings, V. \& Sargent, A.I. 1997, \apj, {\bf 490}, 792.

\reference {} Paresce, F., \& Burrows, C. 1987, \apj, {\bf 319}, L23.

\reference {} P\'erez, M.R., \& Grady, C.A. 1997, Spa.Sci.Rev., {\bf 82}, 407.

\reference {} Pollack, J.B., Hubickyj, O., Bodenheimer, P., Lissaur, J.J., Podolak, M., and Greenzweig, Y. 1996, Icarus, {\bf 124}, 62.

\reference {} Reach, W.T., Abergel, A., Boulanger, F., D\`esert, F.-X., Perault, M., Bernard, J.-P., Blommaert, J., Cesarsky, C., Cesarsky, D., Metcalfe, L., Puget, J.-L., Sibille, F., and Vigroux, L. 1996, A\&A, {\bf 315}, L381.

\reference {} Russell. R.W., Lynch, D.K., Mazuk, A.L., Rossano, G.S., Hanner, M.S., \& Sitko, M.L. 1998, in preparation.

\reference {} Sanford, S.A., \& Walker, R.M. 1985, \apj, {\bf 291}, 838.

\reference {} Sitko, M.L., Grady, C.A., Hanner, M.S., Lynch, D.K., \& Russell, R.W. 1994, in {\it Circumstellar Dust Disks and Planet Formation}, ed. R. Ferlet, and  A. Vidal-Madjar (Paris: Editions Fronti\`eres), p, 389.

\reference {} Sitko, M.L., Lynch, D.K., Russell, R.W., Grady, C.A., \& Hanner, M.S. 1999, in preparation.

\reference {} Stephens, J.R., \& Russell, R.W. 1979, \apj, {\bf 228}, 780.

\reference {} van den Ancker, M.E., Th\'e, P.S., Tjin A Djie, H.R.E., Catala, C., de Winter, D., Blondel, P.F.C., and waters, L.B.F.M. 1997, A\&A, {\bf 324}, L33.

\reference {} Waelkens, C. et al. 1996, A\&A, {\bf 315}, L245.

\reference {} Wetherill, G.W., \& Syewart, G.R. 1993, Icarus, {\bf 106}, 190.

\end {references}


\clearpage


\clearpage

\begin{deluxetable}{ccccc}

\tablewidth{0pc}
\tablecaption{Observing Parameters}
\tablehead{
\colhead{Star}      & \colhead{Mean UT} &
\colhead{Mean Airmass} & \colhead{Number of Spectra} & 
\colhead{Total Integration Time (min)}}
\startdata

HD 163296 & 0540 &  1.98  &  3 & 10.0  \nl
Comet Hale-Bopp & 0610 & 2.08 & 5 & 16.7 \nl
$\alpha$ Lyr & 0645 &   1.92  &  3 & 10.0 \nl
HD 31648  & 1008 &   1.58  &  5 & 16.7 \nl 
$\alpha$ Tau & 0946 &  1.61  &  3 & 10.0 \nl

\enddata
\end{deluxetable}

\clearpage

\begin{figure}
\plotone{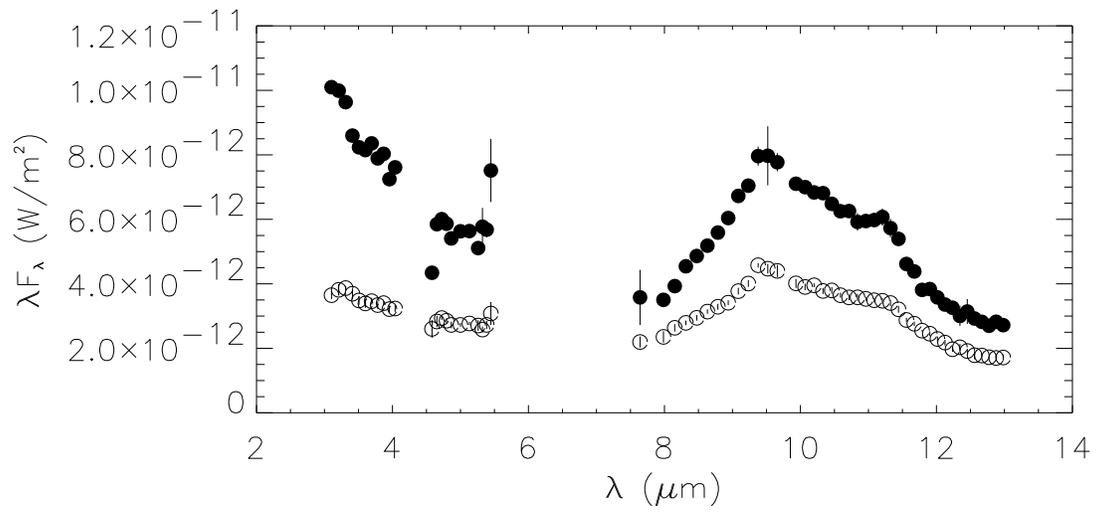}
\figcaption[sitko_fig1.eps]{The flux of HD 31648 (open circles) and HD 163296 filled circles). \label{fig1}}
\end{figure}

\clearpage

\begin{figure}
\plotone{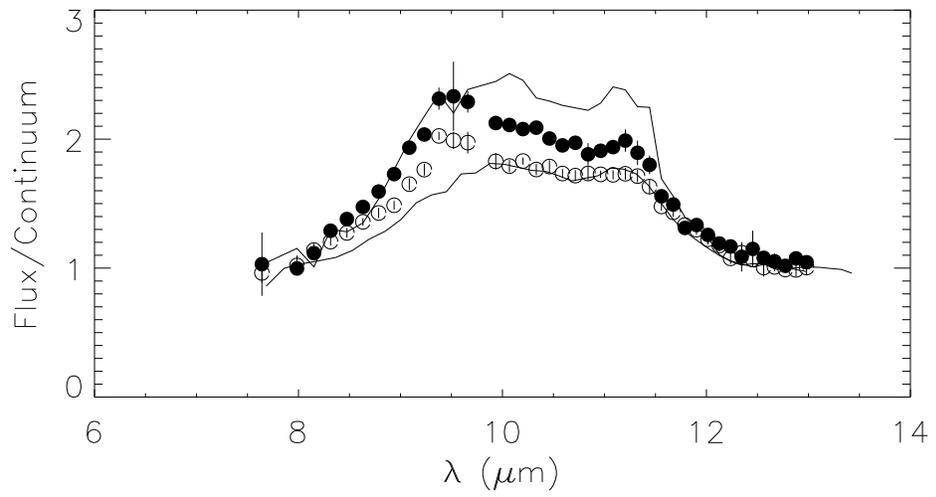}
\figcaption[sitko_fig2.eps]{The flux/continuum ratio for HD 31648 (open circles) and HD 163296 (filled circles) compared to that of Comet Hale-Bopp (upper curve) and Comet Levy (lower curve). \label{fig2}}
\end{figure}

\clearpage

\begin{figure}
\plotone{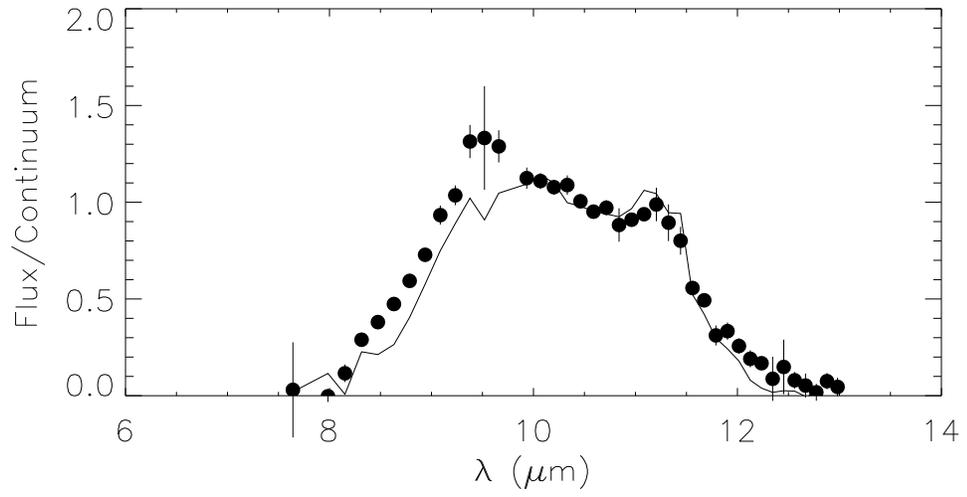}
\figcaption[sitko_fig3.eps]{The spectrum of HD 163296 and Comet Hale-Bopp, after removal of the underlying continua, and scaled to the same approximate strength in the middle of the band. \label{fig3}}
\end{figure}

\clearpage

\begin{figure}
\plotone{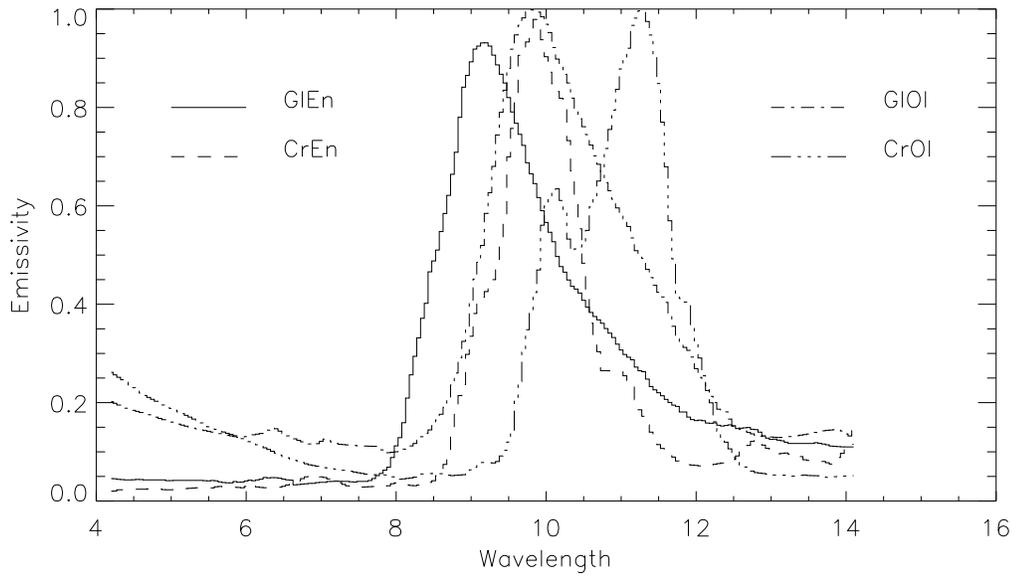}
\figcaption[sitko_fig4.eps]{The emissivities of glassy enstatite (solid line), crystalline enstatite (dashed line), glassy olivine (dash-dot line), and crystalline olivine (dash-dot-dot-dot line), measured in the laboratory by Stephens \& Russell (1979). \label{fig4}}
\end{figure}

\clearpage

\begin{figure}
\plotone{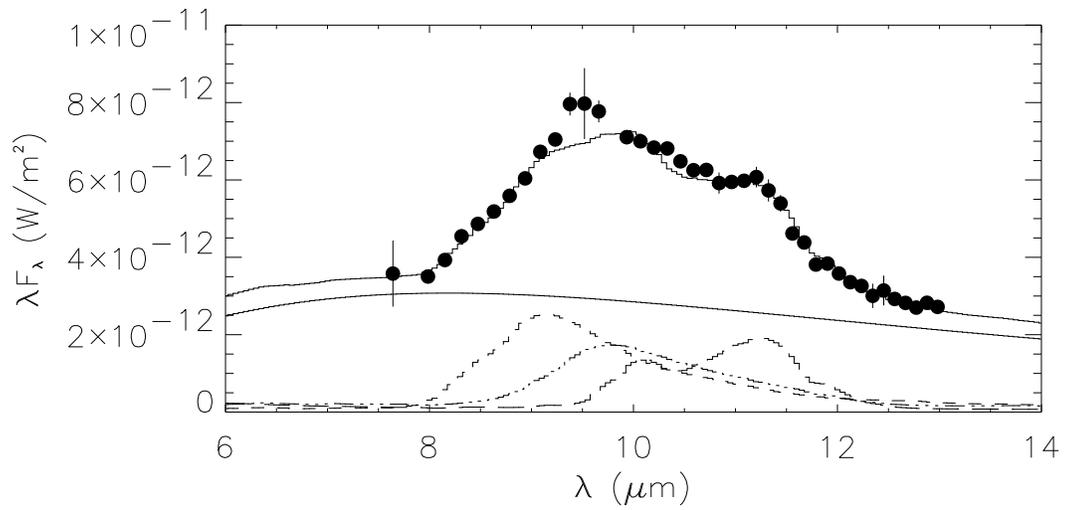}
\figcaption[sitko_fig5.eps]{The spectrum of HD 163296 compared to a model comprised of glassy enstatite, glassy olivine, crystalline olivine, and an underlying blackbody continuum component (all at T=440K). \label{fig5}}
\end{figure}

\end{document}